\documentclass[prl,superscriptaddress,showpacs,twocolumn]{revtex4}
\usepackage{graphicx}

\newcommand\VV{\setbox0=\hbox{V}\hbox{\rm V\raise\ht0
  \hbox to0pt{\hss\vbox to0pt{\hbox{v}\vss}}}}
\def\slashchar#1{\setbox0=\hbox{$#1$}           % set a box for #1
   \dimen0=\wd0                                 % and get its size
   \setbox1=\hbox{/} \dimen1=\wd1               % get size of /
   \ifdim\dimen0>\dimen1                        % #1 is bigger
      \rlap{\hbox to \dimen0{\hfil/\hfil}}      % so center / in box
      #1                                        % and print #1
   \else                                        % / is bigger
      \rlap{\hbox to \dimen1{\hfil$#1$\hfil}}   % so center #1
      /                                         % and print /
   \fi}                                         %
\def\be{\begin{eqnarray}}
\def\ee{\end{eqnarray}}
\newcommand{\ep}{\varepsilon}

\newcommand{\ice}[1]{\relax}

\newcommand{\GeV}{{\rm GeV}}

\newcommand{\nn}{\nonumber}

\begin{document}

\title{$B^0 - \bar B^0$ mixing beyond factorization}

\author{J.G.~K\"orner}

\affiliation{Institut f\"ur Physik, Johannes-Gutenberg-Universit\"at,
Staudinger Weg 7, D-55099 Mainz, Germany}

\author{A.I.~Onishchenko}

\affiliation{Department of Physics and Astronomy,
Wayne State University, Detroit, MI 48201, USA}

\affiliation{Institute of Theoretical and Experimental Physics,
Moscow, 117259 Russia}

\author{A.A.~Petrov}

\affiliation{Department of Physics and Astronomy,
Wayne State University, Detroit, MI 48201, USA}

\author{A.A.~Pivovarov}

\affiliation{Institut f\"ur Physik, Johannes-Gutenberg-Universit\"at,
Staudinger Weg 7, D-55099 Mainz, Germany}

\affiliation{Institute for Nuclear Research of the Russian
Academy of Sciences, Moscow, 117312 Russia}

\begin{abstract}
We present a calculation of the $B^0 -\bar B^0$ mixing matrix element 
in the framework of QCD sum rules for three-point functions. 
We compute $\alpha_s$ corrections to a three-point function at the 
three-loop 
level in QCD perturbation theory, which allows one to extract the matrix 
element with next-to-leading order (NLO) accuracy. This calculation 
is imperative for a consistent evaluation of experimentally-measured 
mixing parameters since the coefficient functions of the effective 
Hamiltonian for $B^0 -\bar B^0$ mixing are known at NLO. 
We find that radiative corrections violate factorization at NLO;
this violation is under full control and amounts to 10\%.
\end{abstract}


\maketitle

The phenomenon of particle-antiparticle mixing, possible in systems of 
neutral mesons of different flavors, is the primary source of studies of 
CP violation~(for review, see e.g.~\cite{buhalla}). 
According to the Cabibbo-Kobayashi-Maskawa (CKM) picture, 
quarks of all three generations must be present in a transition for CP 
violation to occur. Historically, studies of $K^0 - \bar K^0$ mixing provided 
first essential insights into the physics of heavy particles as well as tests of 
general concepts of quantum field theory. For a long time it was the only place 
where the effects of CP violation were clearly established 
(see e.g.~\cite{kkreviewOLD}). 
Since weak couplings of $s$ and $d$ quarks to third generation quarks are
small, experimental studies of CP violation in heavy mesons are considered more 
promising. While recent experimental results for heavy charmed mesons 
$D (\bar u c)$ are encouraging, a full consistent theoretical 
description of this system is still lacking~\cite{Petrov:2002is}. 
These considerations make the systems of $B_d (\bar d b)$ and $B_s (\bar s b)$ 
mesons the most promising laboratory for a precision analysis of CP violation 
and mixing both experimentally and theoretically~\cite{reviewBB}. Hereafter we 
shall consider $B_d$ mesons. The generalization to $B_s$
mesons is straightforward. 

Phenomenologically the system of the $B^0 - \bar B^0$ 
mesons is described by the effective mass operator
$\left(M-i\Gamma/2\right)_{ij}$, $\{i,j\}=\{1,2\}$ which in the presence of 
$\Delta B=2$ interactions acquires non-diagonal terms. The difference 
between the values of the mass eigenstates of $B$ mesons 
$\Delta m = M_{heavy}-M_{light}\approx 2\left|M_{12}\right|$ is an important 
observable which is precisely measured to be
$\Delta m= 0.489\pm0.005(stat)\pm 0.007(syst)~ps^{-1}$~\cite{PDG}.
With an adequate theoretical description, it can be used to extract top quark 
CKM parameters.

In the standard model, the effective low-energy Hamiltonian 
describing $\Delta B = 2$ transitions has been computed at
next-to-leading order (NLO) in QCD perturbation theory (PT)~\cite{Buras}
\begin{eqnarray}
\label{hamilt}
H_{\mbox{eff}}^{\triangle B = 2} &=& \frac{G_F^2M_W^2}{4 \pi^2}
\left({V_{tb}}^{*}V_{td}\right)^2 \eta_B S_0(x_t) \\
&\times&\left[\alpha_s^{(5)}(\mu)\right]^{-6/23} 
\left[1+\frac{\alpha_s^{(5)}(\mu)}{4 \pi} J_5 \right] {\cal O}(\mu) 
\nonumber
\end{eqnarray}
where $\eta_B=0.55\pm0.1$~\cite{Buras:1990fn}, $J_5=1.627$ in the
naive dimensional regularization (NDR) 
scheme, $S_0(x_t)$ is the Inami-Lim function~\cite{Inami:1980fz}, and
${\cal O}(\mu)=(\bar b_L\gamma_{\sigma}d_L)(\bar b_L\gamma_{\sigma}d_L)(\mu)$
is a local four-quark operator at the normalization point~$\mu$. 
Note that the part of Eq.~(\ref{hamilt}) in the second line is
re\-normalization-group (RG) invariant. 
Mass splitting of heavy and light mass eigenstates can then be found
to be
\begin{eqnarray}
\label{offdiag}
&&\Delta m =2 |\langle\bar B^0|H_{\mbox{eff}}^{\triangle B = 2}|B^0
\rangle| \\
&&=~
\!\! {\cal C}\left[\alpha_s^{(5)}(\mu)\right]^{-6/23} 
\left[1+\frac{\alpha_s^{(5)}(\mu)}{4 \pi} J_5 \right] 
\langle\bar B^0|{\cal O}(\mu)|B^0\rangle \nonumber
\end{eqnarray}
where ${\cal C}=G_F^2M_W^2 
\left({V_{tb}}^{*}V_{td}\right)^2 \eta_B m_B S_0(x_t)/\left(4 \pi^2\right)$. 
The largest uncertainty of about 30\% in the theoretical calculation 
is introduced by the poorly known hadronic matrix element
$
{\cal A} = \langle\bar B^0|{\cal O}(\mu)|B^0\rangle
$~\cite{PDG}.
The evaluation of this matrix element is a genuine non-perturbative
task, which can
be approached with several different techniques. The simplest approach 
(``factorization'')~\cite{Gaillard:1974hs} 
reduces the matrix element ${\cal A}$ to the product of matrix
elements measured in leptonic $B$ decays
$
{\cal A}^{f} = (8/3)
$
$
\langle\bar B^0|\bar b_L\gamma_{\sigma}d_L|0\rangle
\langle 0|\bar b_L \gamma^{\sigma}d_L|B^0\rangle = (2/3) f_B^2 m_B^2
$ where the decay constant $f_B$ is defined by
$
\langle 0|\bar b_L \gamma_\mu d_L|B^0({\bf p})\rangle = i p_\mu f_{B}/2
$.
A deviation from the factorization ansatz is usually described by the parameter
$B_B$ defined as
$
{\cal A} = B_B {\cal A}^{f}
$;
in factorization $B_B=1$.
There are many approaches to evaluate this parameter (and the
analogous parameter $B_K$ of $K^0 - \bar K^0$ mixing) available in the
literature~\cite{Bardeen,ope-three-kk,bb-three,Reinders,Narison,Melikhov,lattice,Hiorth}.

The calculation of the hadronic mixing matrix elements 
using Operator Product Expansion (OPE) and QCD 
sum rule techniques for three-point 
functions~\cite{ope-three-kk,bb-three,Reinders} is 
very close in spirit to lattice computations~\cite{lattice}, 
which is a model-independent, first-principles method. 
In the QCD sum rule
approach one relies on asymptotic expansions of a Green's function
while on the lattice the function itself can be computed
in principle. 
The sum rule techniques also provide a consistent way of taking 
into account perturbative corrections to matrix 
elements which is needed to restore the RG invariance of
physical observables usually violated in the factorization
approximation~\cite{kkaplhas}.
The calculation of perturbative corrections to  
$B^0 - \bar B^0$ mixing using OPE and sum rule techniques
is the main subject of this paper. 
A concrete realization of the sum rule method applied here consists of 
the calculation of the moments of the 
three-point correlation function of the interpolating operators of 
the $B$-meson and the local operator ${\cal O}(\mu)$ responsible for 
$B^0 - \bar B^0$ transitions. 

Let us consider the three-point correlation function
\begin{eqnarray}
\label{threepoint1}
\Pi(p_1,p_2)
=\int dx dy 
\langle 0|T J_{\bar B}(x) {\cal O}(0) \bar J_{B}(y)|0\rangle  
e^{i p_2 x - i p_1 y} . \nn
\end{eqnarray}
The operator 
$J_{B} = (m_b+m_d)\bar d i\gamma_5 b$ is chosen 
as interpolating current for the $B^0$-meson and $m_b$ 
is the $b$ quark mass. Note that $J_{B}$ is RG invariant, 
$
J_{B} =\partial_\mu (\bar d \gamma_\mu\gamma_5 b)
$
and 
$
\langle 0|J_{B}(0)|B^0(p)\rangle =  f_{B} m_B^2
$
where $m_B$ is the $B$-meson mass.
A dispersive representation of the correlator reads
\be 
\label{dispdouble}
\Pi(p_1, p_2) \equiv \Pi(p_1^2, p_2^2, q^2)=
\int\frac{\rho(s_1, s_2, q^2)ds_1 ds_2}{(s_1-p_1^2)(s_2-p_2^2)}
\ee
where $q=p_2-p_1$. For the analysis of $B^0 - \bar B^0$ mixing this correlator 
needs to be computed at $q=0$, while within the sum rule framework $q^2=0$. 
This particular kinematical point is infrared safe for massive quarks.
\begin{figure}[t]
\quad\includegraphics[scale=0.3]{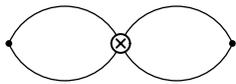}
\caption{Perturbation theory diagram at LO}
\label{figLO}
\end{figure}
The matrix element $\langle\bar B^0| {\cal O}(\mu) |B^0\rangle$
appears in the three-point 
correlator as a contribution of the $B$-mesons in the form of a double pole
\begin{eqnarray}
\label{phenrepr}
\Pi(p_1^2, p_2^2, q^2)=
\frac{\langle J_{\bar B}|\bar B^0\rangle}{m_{B}^2-p_1^2}
\langle\bar B^0|{\cal O}(\mu)|B^0\rangle
\frac{\langle B^0|\bar J_{B}\rangle}{m_{B}^2-p_2^2}+\ldots
\end{eqnarray}
where the ellipsis stand for higher resonances and continuum 
contributions. The matrix element can be extracted by comparing 
the representations given in Eq.~(\ref{phenrepr}) 
and the (smeared) theoretical expression of Eq.~(\ref{dispdouble})
obtained with an asymptotic expansion based on OPE.
Note that 
the analytical calculation of the spectral density itself at NLO of PT
expansion is beyond present computational techniques.
Therefore, a practical way of extracting the $B^0 - \bar B^0$ matrix element
is to analyze the moments of the correlation function at 
${p_1^2=p_2^2=0}$ at the point $q^2=0$. One obtains
\be 
\label{momentsdef}
M(i,j)\equiv 
\frac{\partial^{i+j}\Pi(p_1^2, p_2^2, 0)}
{i!j!\partial p_1^{2i} \partial p_2^{2j}}
=\int \frac{\rho(s_1, s_2, 0)ds_1 ds_2}{s_1^{i+1} s_2^{j+1}}\, .\nn
\ee
A theoretical computation of these moments reduces to
an evaluation of single scale vacuum diagrams 
(we neglect the light quark masses).
This calculation can be done analytically with available tools
for the automatic computation of multi-loop diagrams. 

The leading contribution to the asymptotic expansion is given by 
the diagram shown in Fig.~\ref{figLO}. At leading order (LO) in 
QCD perturbation theory the three-point function of Eq.~(\ref{threepoint1}) 
completely factorizes
$
\Pi(p_1, p_2) = (8/3)$$\Pi_\mu (p_1)\Pi^\mu (p_2)
$
where $\Pi_\mu (p)$ is the two-point correlator
\be 
\label{twopointscorr}
 \Pi_\mu (p)=p_\mu \Pi(p^2)=\int dx e^{i p x}
\langle 0|TJ_{\bar B}(x) \bar b_L \gamma_\mu d_L(0)|0\rangle .
\end{eqnarray}
The calculation of moments is straightforward since
the double spectral density $\rho(s_1, s_2, q^2)$ can be explicitly found.
Using a dispersive representation of $\Pi(p^2)$
\be 
\Pi(p^2) =\int_{m^2}^{\infty} \frac{\rho(s)ds}{s-p^2}, \quad
\rho(s)=\frac{3}{16\pi^2}m^2\left(1-\frac{m^2}{s}\right)^2
\ee
one finds the LO double spectral density
$
\rho^{\rm LO}(s_1,s_2,q^2) =
(8/3) (p_1 \cdot p_2)\rho(s_1)\rho(s_2)=
(4/3)(s_1+ s_2-q^2)\rho(s_1)\rho(s_2)
$.
First {\it non-factorizable} contributions to Eq.~(\ref{dispdouble}) appear
at NLO. Nevertheless, the factorizable diagrams
form an important subset of all contributions, as they are independently gauge
and RG invariant. Thus, a classification of diagrams in terms of their
factorizability is a very powerful technique in the quantitative 
analysis. 

\begin{figure}[t]
\quad\includegraphics[scale=0.3]{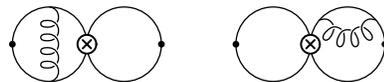}
\caption{Factorizable diagrams at NLO}
\label{figNLOfac}
\end{figure}

The NLO factorizable contributions 
are given by the product of two-point 
correlation functions from Eq.~(\ref{twopointscorr}), 
as shown in Fig.~\ref{figNLOfac}.
Writing 
$
\Pi(p^2)=\Pi_{\rm LO}(p^2)+\Pi_{\rm NLO}(p^2)
$
we obtain
$\Pi_{\rm NLO}^f(p_1, p_2)$$ = $$(8/3)$$(p_1.p_2)$$
(\Pi_{\rm LO}(p_1^2)\Pi_{\rm NLO}(p_2^2)
$
$
+$$\Pi_{\rm NLO}(p_1^2)\Pi_{\rm LO}(p_2^2))
$.
The spectral density of the correlator
$\Pi_{\rm NLO}(p^2)$ is known analytically.
This completely solves the problem of the NLO analysis in 
factorization. Note that even a NNLO analysis of factorizable 
diagrams is possible as several moments of two-point correlators are 
known analytically. Others can be obtained numerically 
from the approximate spectral density~\cite{chetmomnondiag}. 

\begin{figure}[t]
\quad\includegraphics[scale=0.3]{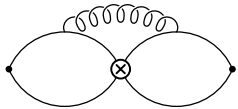}
\caption{An example of a non-factorizable diagram at NLO}
\label{figNLOnonfac}
\end{figure}

The NLO analysis of non-factorizable contributions within perturbation
theory is the main result of 
this paper. This analysis amounts to the calculation of a set of 
three-loop diagrams (a typical diagram is presented in Fig.~\ref{figNLOnonfac}). 
These diagrams can be computed using the package MATAD for automatic calculation
of Feynman diagrams~\cite{matad}. Before applying this package,
the combinatorics of disentangling the tensorial structures 
has to be solved and all the diagrams have to be reduced to 
a set of scalar integrals which can be done using the results 
of ref.~\cite{Davyd}. The steps described above were automized with the
computer algebra system FORM~\cite{form}. We shall present the details of this calculation 
elsewhere.

The local four-quark operator ${\cal O}$ entering the 
effective Hamiltonian has to be renormalized. 
We employ dimensional regularization with an anticommuting $\gamma_5$.
The renormalization of the operator ${\cal O}$ reads
\be
{\cal O}^R={\cal O}^B-\frac{\alpha_s}{4\pi}\frac{1}{\ep}{\cal O}_c
\ee
with
$
{\cal O}_c=(\bar b_L \Gamma_{\mu\nu\alpha} t^a d_L)
(\bar b_L \Gamma^{\mu\nu\alpha} t^a d_L)
$. The $t^a$ are the $SU_c(3)$ ge\-nerators 
and 
$
\Gamma_{\mu\nu\alpha}$$=(\gamma_\mu\gamma_\nu\gamma_\alpha$$-
\gamma_\alpha\gamma_\nu\gamma_\mu)/2$.
The renormalization of the factorizable contributions reduces 
to that of the $b$-quark mass $m$. 
We use the quark pole mass as a mass 
parameter of the calculation.

The expression for the ``theoretical'' moments reads
\be
\label{thfull}
M_{th}(i,j)=\frac{m^6 a_{ij}}{m^{2(i+j)}}
\left(1+\frac{\alpha_s}{4\pi} \left(b^{f}_{ij}+b^{nf}_{ij}\right)\right)
\ee
where the quantities $a_{ij}$, $b^{f}_{ij}$ and $b^{nf}_{ij}$
represent LO, NLO factorizable and NLO nonfactorizable contributions 
as shown in Figs.~\ref{figLO}-\ref{figNLOnonfac}. 
The NLO nonfactorizable contributions $b^{nf}_{ij}$ with 
$i+j\leq 7$
are analytically calculated in this paper for the first time.
The calculation required about 24 hours of computing time on a dual-CPU 
2 GHz Intel Xeon machine. The calculation of higher moments is feasible 
but requires considerable optimization of the code. This work
is in progress and will be presented elsewhere.
As an example, we give the analytical results for
the lowest finite moment $M_{th}(2,2)$:
\be
a_{22}=\frac{1}{(16\pi^2)^2}
\left(\frac{8}{3}\right), \quad
b^{f}_{22}=\frac{40}{3}+\frac{16\pi^2}{9}\, ,
\ee
\be
b^{nf}_{22}
=
S_2\frac{8366187}{17500}-\zeta_3\frac{84608}{875}
-\pi^2\frac{33197}{52500}
-\frac{426319}{315000}\nn .
\ee
Here
$
S_2=\frac{4}{9\sqrt{3}}{\rm Cl}_2
\left(\frac{\pi}{3}\right)=0.2604\ldots
$, $\zeta_3=\zeta(3)$, and $\mu^2=m^2$. 
For higher moments we present only numerical values 
for 
$
b^{nf}_{ij}
$:
$
b^{nf}_{2(2345)}=\{0.68,1.22,1.44,1.56\}
$
and 
$
b^{nf}_{3(34)}=\{1.96,2.25\}
$.
The non\-perturbative contribution due to the gluon condensate
is small. For the standard numerical 
value $\langle g^2_s G^2\rangle=0.5~\GeV^4$~\cite{svz},
the nonfactorizable contribution due to the gluon condensate 
amounts to about 3\% of the 
perturbative contribution for the high moment $M(3,3)$.
For lower moments it is even smaller and, therefore,
can be safely neglected in the whole analysis.

We use the above theoretical results to analyze sum rules 
and extract the non-perturbative parameter $B_B$. 

The ``phenomenological'' side of the sum rules is given by the moments
which can be inferred from Eq.~(\ref{phenrepr})
\begin{eqnarray}
\label{phenfull}
&&M_{ph}(i,j)
= \frac{8}{3} B_B
\frac{f_B^4 m_B^2 }{m_B^{2(i+j)}}\\
&&+{\cal O}\left(\frac{1}{(m_B^2+\Delta)^{i+1}m_B^{2j}},
\frac{1}{(m_B^2+\Delta)^{j+1}m_B^{2i}}\right)\nn
\end{eqnarray}
where the contribution of the $B$-meson is displayed
explicitly. The remaining parts are the contributions due to higher resonances and 
the continuum which are suppressed due to the mass gap $\Delta$ in the 
spectrum model.

For comparison we consider the factorizable approximation
for both ``theoretical''
\be
\label{thfact}
M_{th}^{f}(i,j)=\frac{m^6 a_{ij} }{m^{2(i+j)}}\left(1
+\frac{\alpha_s}{4\pi} b^{f}_{ij}\right)
\ee
and ``phenomenological'' moments, which, by construction,
are built from the moments of the two-point function
of Eq.~(\ref{twopointscorr})
\be
\label{phenfactprod}
M_{ph}^{f}(i,j)
= \frac{8}{3}\frac{f_B^4 m_B^2}{m_B^{2(i+j)}}+...
\end{eqnarray}
According to standard QCD sum rule technique, the 
``theoretical'' calculation is dual to the ``phenomenological''
one. Thus, Eq.~(\ref{phenfull}) should be equivalent 
(in the sum rule sense) to Eq.~(\ref{thfull}). Also, in factorization,
Eq.~(\ref{phenfactprod}) is equivalent  to Eq.~(\ref{thfact}). 
Now Eq.~(\ref{thfact}) and
Eq.~(\ref{thfull}) differ only due to non-factorizable corrections. Therefore, 
the difference between Eq.~(\ref{phenfactprod}) and Eq.~(\ref{phenfull}) is
because the residues differ from their factorized values.

To find the nonfactorizable addition to $B_B$ from the sum rules 
we form ratios of the total and factorizable contributions.
On the ``theoretical'' side one finds
\be
\label{fintheory}
\frac{M_{th}(i,j)}{M_{th}^{f}(i,j)}
=1+\frac{\alpha_s}{4\pi}\frac{b^{nf}_{ij}}{1+\frac{\alpha_s}{4\pi}
b^{f}_{ij}}\, .
\ee
This ratio is mass-independent.
On the ``phenomenological'' side we have
\be
\frac{M_{ph}(i,j)}{M_{ph}^{f}(i,j)}=
\frac{B_B+R_B (z^j+z^i)+C_Bz^{i+j}}
{1+R^f(z^j+z^i)+C^f z^{i+j}}
\ee
where $z=m_B^2/(m_B^2+\Delta)$ is a parameter that describes the
suppression of higher state contributions. $\Delta$ is a gap 
between the squared masses of the $B$-meson and higher states. 
$R_B$, $C_B$, $R^f$ and $C^f$ 
are parameters of the model for higher state contributions within the sum
rule approach. 
In order to extract the non-factorizable
contribution to $B_B$ we write $B_B=1+\Delta B$.
Similarly, one can parameterize contributions to ``phenomenological'' moments 
due to higher $B$-meson states by writing $R_B=R^f+\Delta R$ and 
$C_B=C^f+\Delta C$. Clearly, $\Delta B=\Delta R=\Delta C=0$ in factorization. 
We obtain
\be
\label{finphen}
\frac{M_{ph}(i,j)}{M_{ph}^{f}(i,j)}=1
+\frac{\Delta B+\Delta R (z^j+z^i)+\Delta C z^{i+j}}
{1+R^f(z^j+z^i)+C^f z^{i+j}}\, .
\ee
Comparing Eqs. (\ref{fintheory}) and (\ref{finphen}) one sees 
how the perturbative non-factorizable correction $b^{nf}_{ij}$
is ``distributed'' 
among the phenomenological parameters of the spectrum. 
We extract $\Delta B$
by a combined fit of several ``theoretical'' and ``phenomenological'' moments.
The final formula for the determination of $\Delta B$ reads
\be
\frac{\alpha_s}{4\pi} b^{nf}_{ij}
=\Delta B
+\Delta R (z^{j-2}+z^{i-2})
+\Delta C z^{i+j-4}
\ee
where $\Delta R$ and $\Delta C$ are free parameters of the fit.
We take $\Delta=0.4 m_B^2$ which corresponds to the duality interval of
$1~{\rm GeV}$ in energy scale within the HQET analysis of the $B$ meson 
two-point correlator. We then perform least-squares fit to determine 
$\Delta B$. 
Using all available theoretical moments we find 
$(\Delta B,\Delta R,\Delta C)$$=\alpha_s(m)/(4\pi)$$(7.1,-5.0,3.6)$.
We checked the stability of the sum rules which lead 
to a prediction of $\Delta B$.
It can be illustrated in the following way. The contribution of higher
$B$ states is suppressed more strongly for higher moments and therefore  
decreases with increasing order of a moment, while the perturbative
correction grows. The sum of both is (approximately) the same for all 
moments, which leads to a (almost) constant value for $\Delta B$,
independent of the particular moment.
The calculation can be further improved with the evaluation of 
higher moments. The result is sensitive to the parameter $z$ or to
the magnitude of the mass gap $\Delta$ used in the parametrization of the
spectrum. Estimating all uncertainties we
finally find the NLO non-factorizable QCD corrections to
$\Delta B$ due to perturbative contributions to the sum rules to be
$
\Delta B=(6\pm 1)\alpha_s(m)/(4\pi)
$.
For $m=4.8~{\rm GeV}$, $\alpha_s(m)=0.2$~\cite{PDG,Penin:1999kx} it leads to 
$
\Delta B=0.095\approx 0.1
$.
It is known that nonperturbative corrections (such as the ones due to the quark-gluon 
condensate) to the parameter $B_B$ are negative, 
$
\Delta B^{nonPT}(m)=-0.05
$~\cite{bb-three}.
Combining this result with the present analysis we find
$
B_B(m)=1+0.1_{PT}-0.05_{nonPT}
$
showing the excellent numerical validity of the factorization
approximation at the scale $\mu=m$.

In conclusion, we have evaluated the $B^0 -\bar B^0$ mixing matrix element 
in the framework of QCD sum rules for three-point functions at NLO
in perturbative QCD. The effect of radiative corrections on $B_B$ is under 
complete control and amounts to approximately $+10$\%. We have also
shown that
perturbative QCD correction to $\Delta B$ for the moments considered
in our analysis completely dominates the correction due to the
gluon condensate.

We thank Rob Harr for fruitful discussions.  
This work was supported in part by
the Russian Fund for Basic
Research under contracts 01-02-16171 and 02-01-00601, 
by Volkswagen grant No. I/77788, and by the US Department of Energy under 
grant DE-FG02-96ER41005.

\end{document}